\documentclass[12pt]{article}%
\usepackage[utf8]{inputenc}%
\usepackage[english]{babel}%
\usepackage[dvips]{graphicx}%
\usepackage{hhline}%
\usepackage{longtable}%
\begin{document}

\begin{center}
{\bf \large ON THE KINEMATICS OF THE LOCAL COSMIC VOID}\\
\bigskip
O.\,G.\,Nasonova and I.\,D.\,Karachentsev\\
\bigskip
{\footnotesize{}Special Astrophysical Observatory of the Russian Academy of Sciences,\\
Nizhnij Arkhyz, KChR, 369167, Russia}\\
\bigskip
\end{center}

{\itshape{} We collected the existing data on the distances and radial
velocities of galaxies around the Local Void in the Aquila/Hercules to examine
the peculiar velocity field induced by its underdensity. A sample of 1056
galaxies with distances measured from the Tip of the Red Giant Branch, the
Cepheid luminosity, the SNIa luminosity, the surface brightness fluctuation
method, and the Tully-Fisher relation has been used for this purpose. The
amplitude of outflow is found to be $\sim$300~km$\:$s$^{-1}$. The galaxies
located within the void produce the mean intra-void number density about 1/5 of
the mean external number density of galaxies. The void's population has a lower
luminosity and a later morphological type with the medians: $M_B = -15.7^m$ and
T = 8 (Sdm), respectively.}

{\bf{}Keywords}: Galaxies ---  large-scale structure of the Universe.

\section{Introduction}

Since the cosmic voids were first detected (Joveer et al. 1978, Gregory \&
Tompson 1978, Kirshner et al. 1981, de Laparent et al. 1986), it gradually
became clear that numerous empty volumes with dimensions of $\sim10-50$~Mpc are
the main architectural elements of the large-scale structure of the Universe
(van de Weygaert \& Platen, 2009). On the example of the Local Volume
($D<10$~Mpc) it was shown (Tikhonov \& Karachentsev 2006) that there also exist
smaller mini-voids and bubbles with a diameter of $\sim1-5$~Mpc. The closest of
the existing cosmic lacunae was detected by Tully \& Fisher (1987) while
compiling the Atlas of nearby galaxies. Its central part is located in the
constellations of Aquila and Hercules, in the region of strong Galactic
absorption. Nevertheless, an apparent deficit of galaxies in this region is only
partially due to the absorption of light. The surveys of the Local Void in the
21 cm neutral hydrogen line, performed in the Parkes and Arecibo observatories
(Zwaan et al. 2005, Kraan-Korteweg et al. 2008, Giovanelli et al. 2005,
Saintonge et al. 2008) have confirmed low local density of galaxies with radial
velocities below 3000~km$\:$s$^{-1}$. Identifications of the IRAS sources
accompanied by measurements of their radial velocities (Roman et al. 1996,
Nakanishi et al. 1997), as well as the search for dwarf galaxies of low surface
brightness (Karachentseva et al. 1998) did not eliminate the observed underrun
of nearby galaxies in the region of the Local Void, which occupies about 1/6 of
the entire sky.

The dimensions of the Local Void in depth and the extent of its galaxy
population are still debated. Tully et al. (2008) note that the region of low
density extends up to distances of $\sim(40-60)$~Mpc. Kraan-Korteweg et al.
(2008) suggest that the local region of depression may be even larger,
neighboring a more distant void in Microscopium/Sagittarius. Furthermore, these
authors denote the existence of a few filaments inside this volume, dividing the
``supervoid'' into 2 or 3 voids with the size of 10--30 Mpc.

This whole extended region with a low density of galaxies is located
approximately along the north pole of the supergalactic coordinate system, that
intensifies the concentration of nearby galaxies towards the Local Supercluster
plane. According to results of numerical simulations in the $\Lambda$CDM models
(Ceccarelli et al. 2008, Schaap 2007, Tully et al. 2008), the expansion of
cosmic voids occurs more rapidly and is characterized by a local excess of the
Hubble constant $\Delta H\simeq 0.2H_0$. At the typical void radius of
$\sim15$~Mpc, peculiar velocities of galaxies revealed at its edges amount to
around $\pm$ (220)~km$\:$s$^{-1}$. As Tully et al. (2008) argue, the presence of
a vast Local Void in the direction of $+$SGZ generates the velocity of
($V_{pec})_{LG}\simeq 260$~km$\:$s$^{-1}$ of the Local Group relative to the
cosmic microwave background. The direction of this component towards $-$SGZ
explains a well-known phenomenon of the ``Local Velocity Anomaly'' (Faber \&
Burstein 1988), which remained a mystery during the past 20 years.

The observational data (Karachentseva et al. 1999, Kraan-Korteweg et al. 2008)
show that the Aquila/Hercules region with the coordinates RA=[17.0$^h - 21.0^h$]
and Dec=[$-30^{\circ}, +40^{\circ}$] reveals an almost complete lack of galaxies
with radial velocities below 1500~km$\:$s$^{-1}$. This region of the sky is
demonstrated in Fig.$\;$1, where the shaggy diagonal band marks the region of
strong absorption $A_B>2.0^m$ according to Schlegel et al. (1998). The galaxies
with radial velocities $V_{LG}<1500$~km$\:$s$^{-1}$ in the Local Group rest
frame are marked with circles, while the figures indicate the radial velocity of
a given galaxy. In the velocity interval of [1600--2200]~km$\:$s$^{-1}$ in this
region there appear the galaxies that are likely to belong to the far wall of
the nearby void, or to the filament that splits a larger void into sectors.

Judging from the distribution of galaxies in Fig.$\;$1, the geometric center of
the Local Void is located near RA=19.0$^h$, Dec=$+3^{\circ}$, which practically
coincides with the position of its center according to Tikhonov \& Karachentsev
(2006). Note that other estimates appeared in the literature: RA=18.8$^h$,
Dec=$-3^{\circ}$ (Tully et al. 2008), and RA=18.6$^h$, Dec=$+18^{\circ}$
(Karachentseva et al. 1999).

\begin{figure}[h]
\begin{center}
\includegraphics[height=0.8\textwidth,keepaspectratio,angle=270]{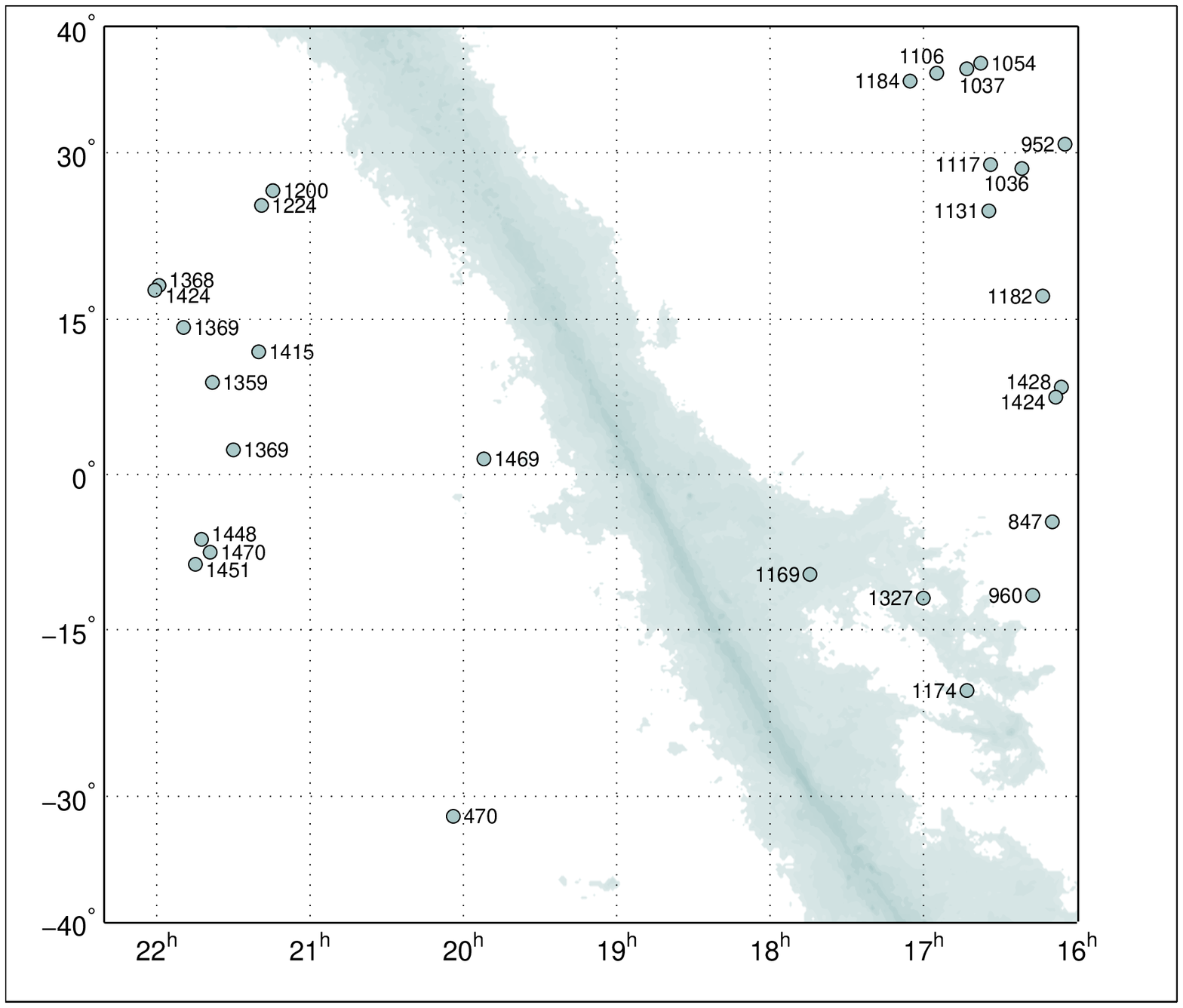}
\vbox{\vspace{0.6cm}
\parbox{0.8\textwidth}{\footnotesize{}%
Fig.$\;$1. Distribution of galaxies in the sky in the Local Void region in
equatorial coordinates. The numbers indicate the radial velocity of galaxies
relative to the Local Group centroid. The diffuse band depicts the zone of
strong galactic absorption.}}
\end{center}
\end{figure}

Below we consider the structure and kinematics of the Local Void as the near
part of a more extended depression among the galaxies observed along the north
pole of the Local Supercluster. To be definite, we assume that the Local Void
begins right at the edge of the Local Group and extends approximately 20 Mpc,
with the center at RA=19.0$^h$, Dec=$+3^{\circ}$ at the distance of
$D_{LV}=10.0$~Mpc from the observer.

\section{Observational data}

In order to analyze the peculiar velocity field in the vicinity of the Local
Void, we collected a sample of galaxies with known radial velocities and
distances that are located within a spherical volume of radius $R_{LV}=25$~Mpc
around the void center fixed above. Such a sufficiently large radius, 2.5 times
the assumed radius of the void, was selected to have a representative
asymptotics of the velocity field around the void. The NED and HyperLeda
databases were our sources of data on the radial velocities of galaxies. Radial
velocities were expressed relative to the centroid of the Local Group, with the
apex parameters used in the NED.

The distances to the galaxies in our sample were measured in different ways. For
nearby galaxies, the most versatile and accurate method of distance estimation
is to use the luminosity of the Tip of the Red Giant Branch (TRGB). The
application of this method, proposed by Lee et al. (1993), to the images of
galaxies, obtained with the WFPC2 and ACS cameras of the HST, gave accurate
distances to more than 250 galaxies in the Local Volume. A summary of TRGB
distances is contained in the Catalog of Neighboring Galaxies (=CNG,Karachentsev
et al. 2004). Using the CNG, we have adjoined more recent estimates of distances
(Karachentsev et al. 2006, Tully et al. 2006), as well adding to them a number
of galaxies with distances measured from the luminosity of Cepheids.

For the early type (E,S0) galaxies dominated by old stellar population, the most
effective method to measure distances is based on the fluctuations of surface
brightness. Tonry et al. (2001) measured by this method the distances to 300 E
and S0 galaxies with typical velocities $cz<4000$~km$\:$s$^{-1}$.

We have as well included in our sample a small number of galaxies with
high-precision distance measurements from the luminosity of SNIa type Supernovae
(Tonry et al. 2003).

Kashibadze (2008) used a multiparametric near-infrared Tully-Fisher relation to
determine distances to $\sim$400 spiral galaxies from the 2MASS Selected Flat
Galaxy Catalog by Mitronova et al. (2004) with radial velocities
$cz<3000$~km$\:$s$^{-1}$. The zero-point of this NIR-TF relation was calibrated
by 15 galaxies with distance estimated from the Cepheids and TRGB.

Finally, we used compilations of distance estimates based on the classic optical
(one-parametric) Tully-Fisher relations in different photometric bands (B, R,
I), presented by Tully et al. (2008, 2009) and Springob et al. (2007). To
calibrate these data we used the galaxies with distances measured via the TRGB
and Cepheids.

In total our sample contains 1056 galaxies. The distribution of their number by
the method of distance estimation is presented in Table~1. The last two columns
of the Table indicate the distance modulus measurement error, typical for each
method, and a statistical significance of each subsample i.e. its ``goodness'',
$G=(N/100)^{1/2}\cdot\sigma^{-1}_m$.

\begin{table}
\caption{\footnotesize{}Number of galaxies around the Local Void with distances
measured by different methods.}
\begin{center}
{\footnotesize{}%
\begin{tabular}{lrcr} \hline
    Sample     & Number  & $\sigma_m$ &  Goodness   \\
\hline
   TRGB+Ceph   &   264   &   0.15  &    10.8     \\
   SBF (Tonry) &   141   &   0.25  &     4.7     \\
   SNIa (Tonry)&    11   &   0.10  &     3.3     \\
   TF (NIR)    &   139   &   0.40  &     2.9     \\
   TF (opt)    &   501   &   0.40  &     5.6     \\
\hline
    All        &  1056   &    -    &    27.3     \\
\hline
\end{tabular}}
\end{center}
\end{table}

\begin{figure}[p]
\begin{center}
\includegraphics[height=1\textwidth,keepaspectratio,angle=270]{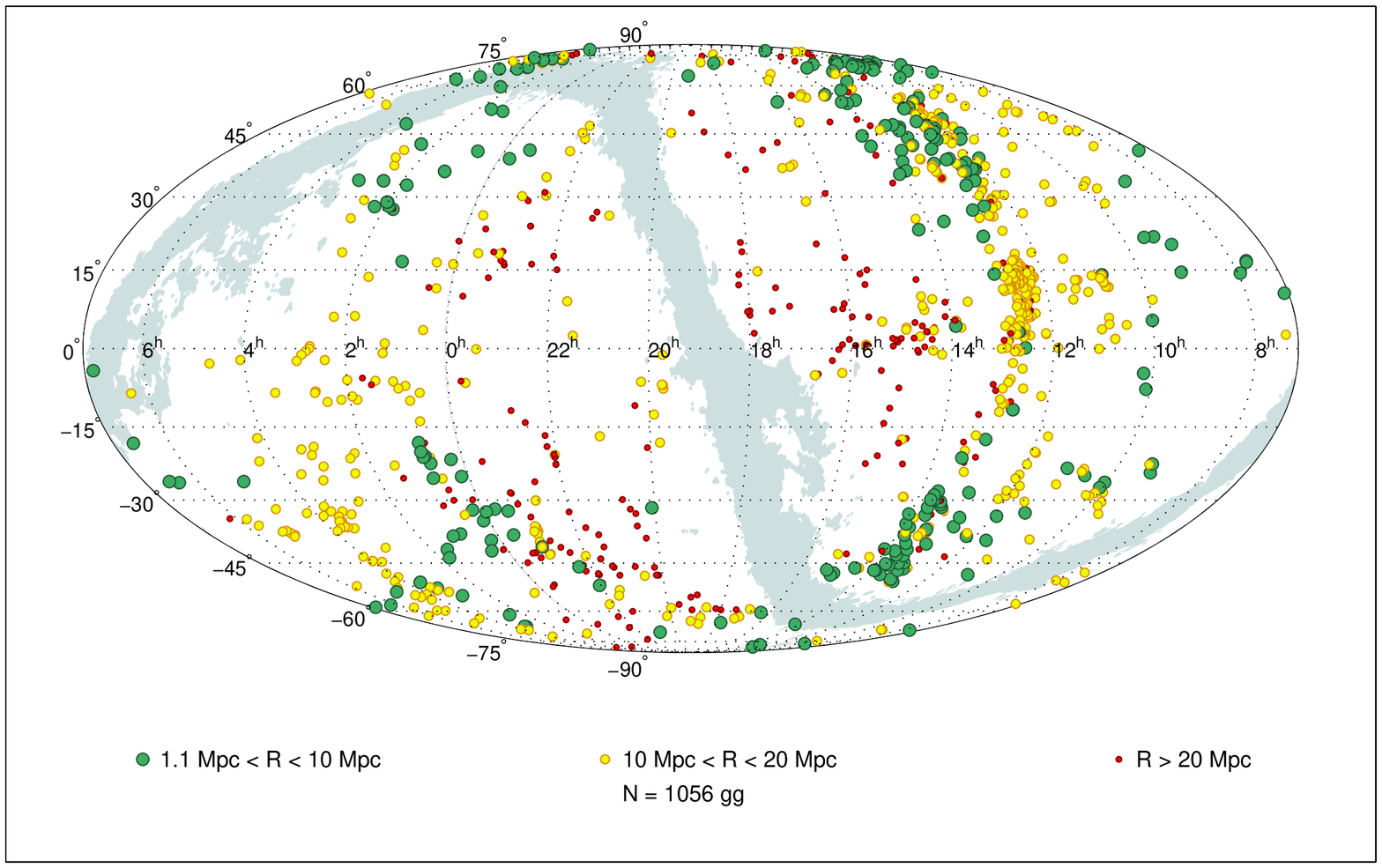}
\vbox{\vspace{0.6cm}
\parbox{0.8\textwidth}{\footnotesize{}%
Fig.$\;$2. Distribution of 1056 galaxies in the sky with distances from the
center of the Local Void within 25 Mpc. The galaxies in three intervals of
distance from the observer are marked with circles of different diameters.}}\par
\vspace{1.2cm}
\includegraphics[height=0.8\textwidth,keepaspectratio,angle=270]{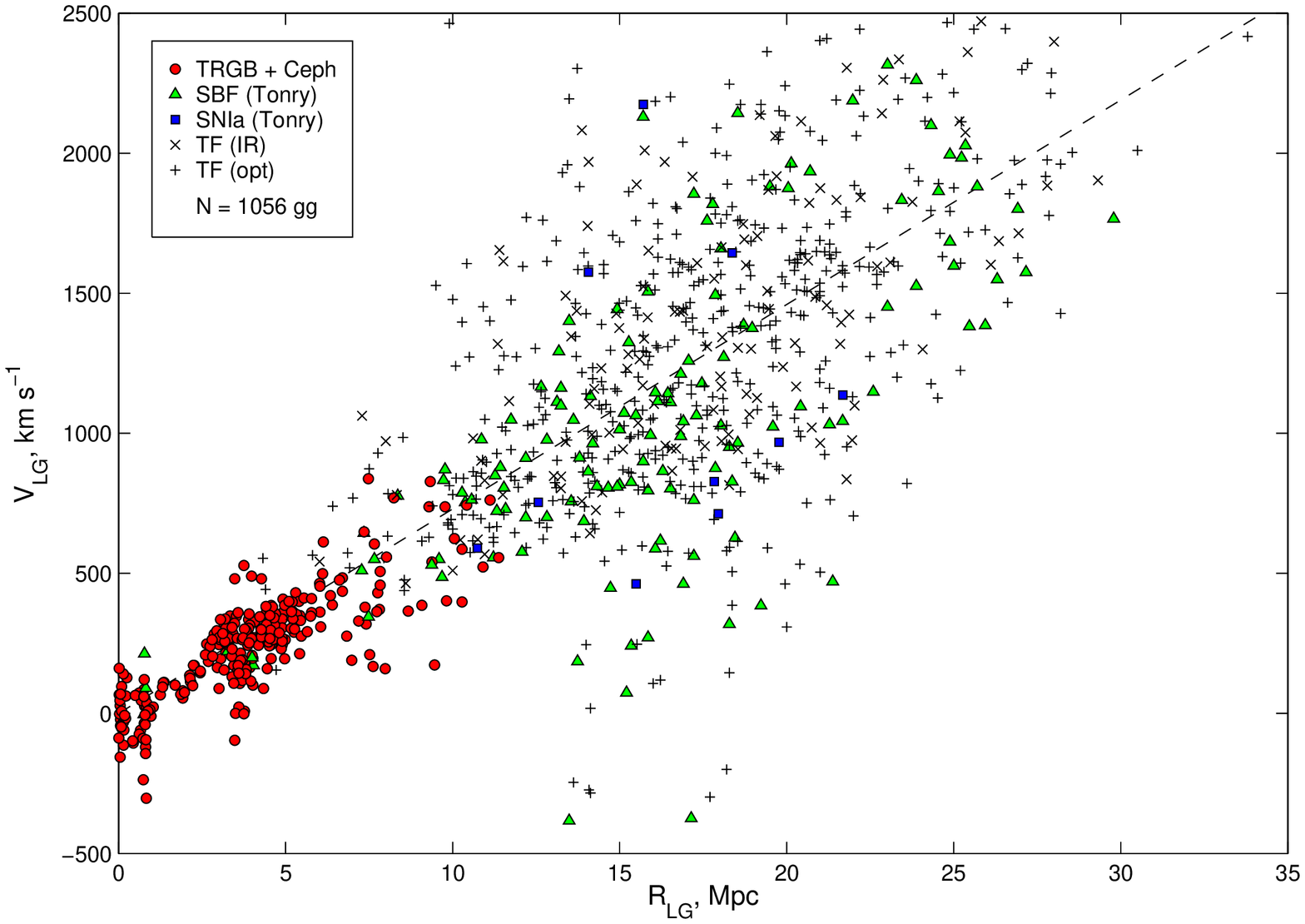}
\vbox{\vspace{0.4cm}
\parbox{0.8\textwidth}{\footnotesize{}%
Fig.$\;$3. Distribution of 1056 galaxies by radial velocities and distances from
the observer. Galaxies with different sources of distances are marked with
different symbols.}}
\end{center}
\end{figure}

Figure~2 shows the distribution in the sky of 1056 galaxies from our sample in
equatorial coordinates. Galaxies with close ($R<10$~Mpc), medium (10--20 Mpc)
and far ($R>20$~Mpc) distances are shown by circles of different diameters. The
region of strong absorption ($A_B>2.0^m$) is filled with gray. The members of
the Local Group with $R<1.1$~Mpc were excluded. The Hubble relation between
radial velocities and distances of 1056 galaxies relative to the observer is
presented in Figure~3. The galaxies with distance estimates made using different
methods are depicted by different symbols. The near part of the Hubble diagram
($R_{LG}<10$~Mpc) is dominated by the galaxies with high-accuracy distance
measurements using the TRGB and Cepheids, and at larger distances their
estimates are derived mainly from the Tully-Fisher diagrams, and the
fluctuations of surface brightness. An increased scatter of galaxies relative to
the line with an inclination of $H_0=73$~(km$\:$s$^{-1}$)/Mpc (the dashed line)
at $R_{LG}\simeq 15-20$~Mpc exists owing to the virial velocities of members of
the Virgo cluster.

\section{Discussion}

When there exists an expansion of the void's neighborhood, then the galaxies,
located for the observer in front of the empty volume, should have the
velocities systematically lower than the Hubble velocity, and the objects behind
the far edge of the void shall have radial velocities higher than the Hubble
velocity. The pattern of deviations from the pure Hubble flow will be
characterized by a wave with an amplitude of an opposite sign to that, observed
in the pattern of the Virgo-centric flow and Fornax-centric flow around these
nearby attractors (Karachentsev \& Nasonova 2010, Nasonova et al. 2011). Here
the amplitude of outflow velocity from the void is expected to be the lower, the
further the angular distance of a galaxy $\Theta$ from the the void center.

\begin{figure}[p]
\begin{center}
\includegraphics[height=0.8\textwidth,keepaspectratio,angle=270]{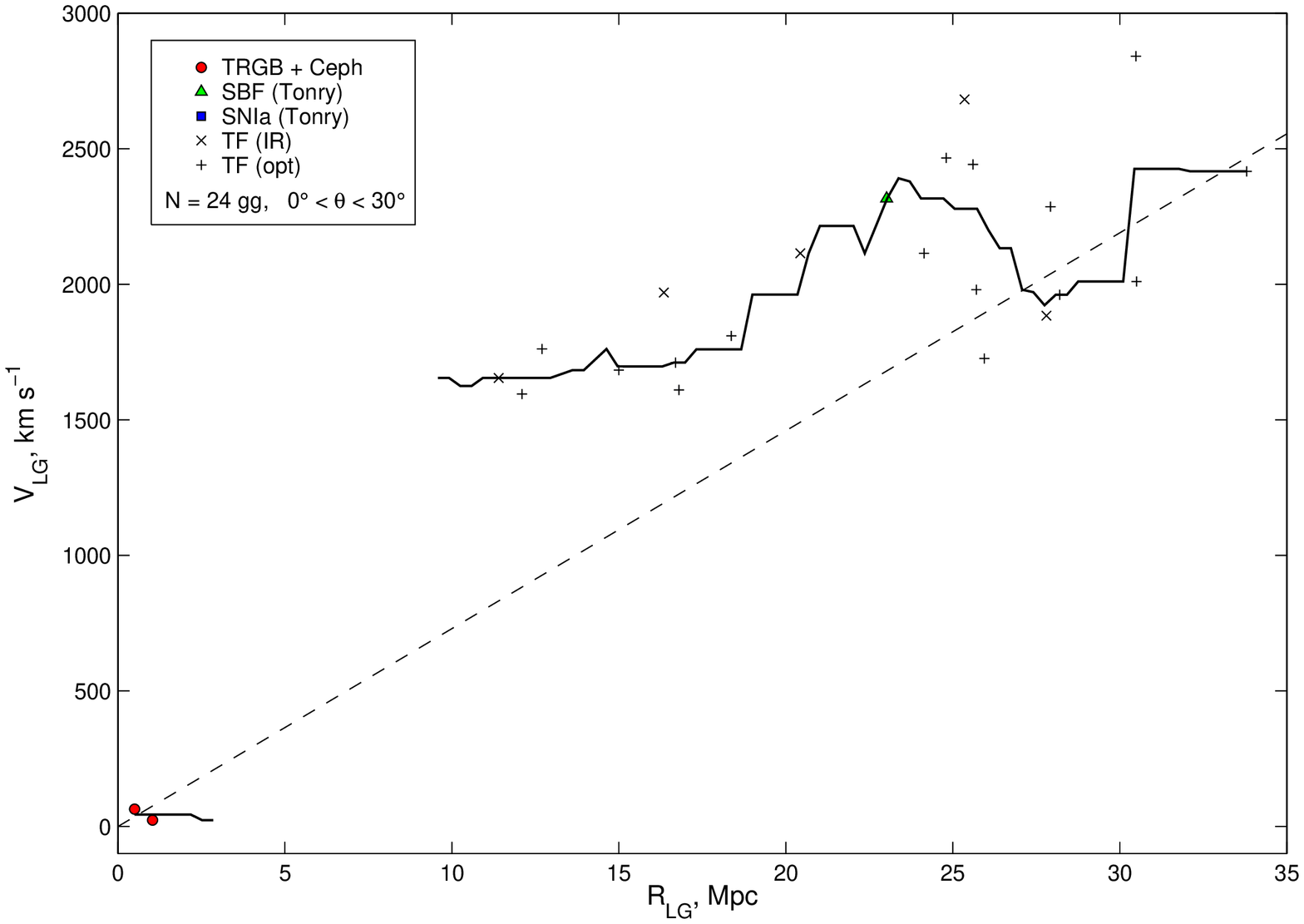}
\includegraphics[height=0.8\textwidth,keepaspectratio,angle=270]{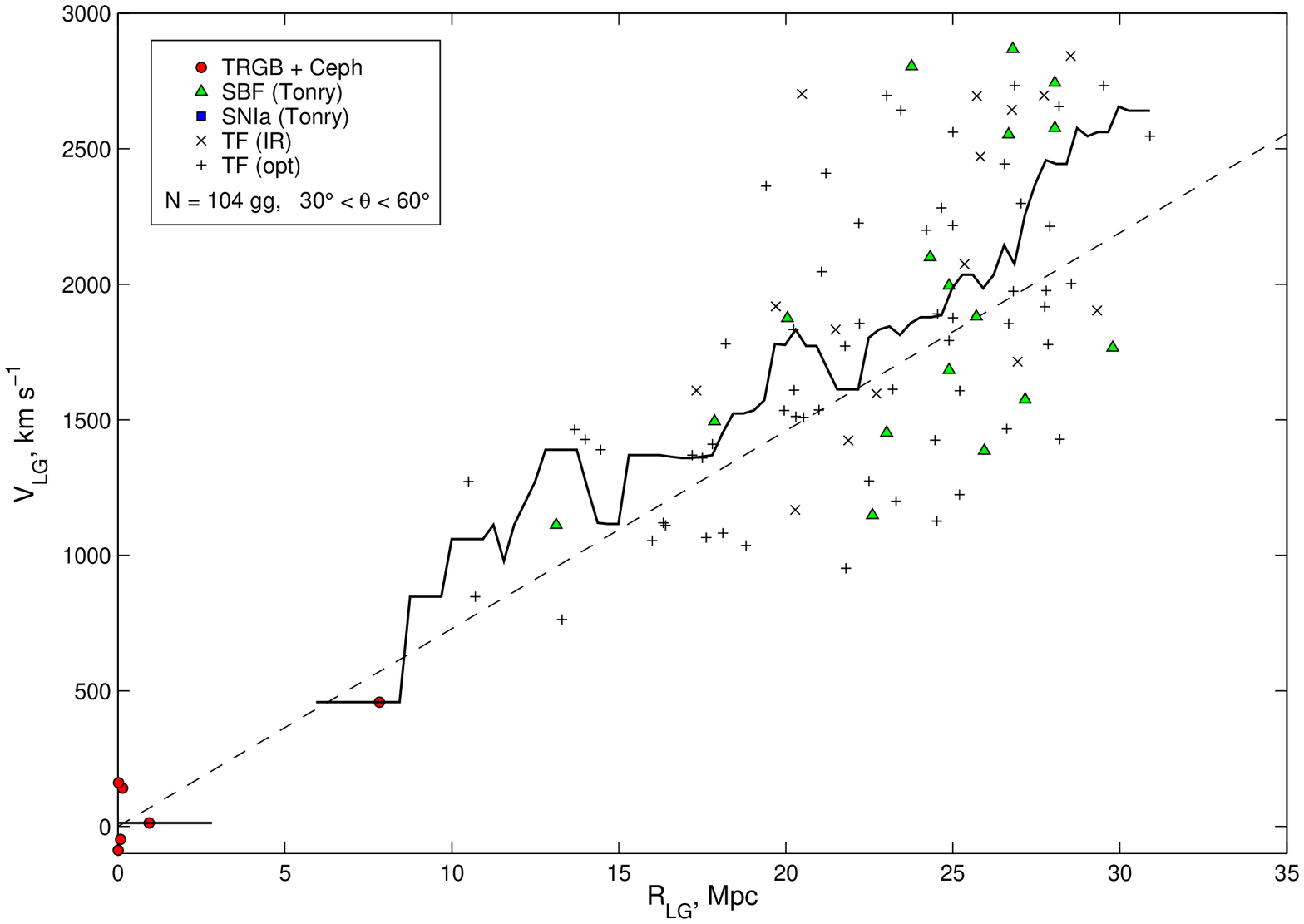}
\vbox{\vspace{0.4cm}
\parbox{0.8\textwidth}{\footnotesize{}%
Fig.$\;$4. Hubble diagram for galaxies with angular distances from the Local
Void center below $30^{\circ}$ (top panel) and $30-60^{\circ}$ (bottom panel).}}
\end{center}
\end{figure}

The top and bottom panels in Fig.$\;$4 reproduce the Hubble diagrams for 24 and
104 galaxies, located in the cones of $\Theta<30^{\circ}$ and
$30^{\circ}<\Theta<60^{\circ}$, respectively. Since the Local Void begins
literally at the doorstep of the Local Group, we can not actually detect the
nearby part of the ``outflow wave''. Some of the closest galaxies at
$R_{LG}\simeq 1$~Mpc are the members of the Local Group. On the far side behind
the void, in the $\Theta<30^{\circ}$ cone the anticipated excess of radial
velocities of galaxies is observed. The broken line in the figure represents a
sliding median with the averaging window of 2 Mpc. Its excess over the Hubble
line at $H_0$=73~(km$\:$s$^{-1}$)/Mpc reaches $\sim500$~km$\:$s$^{-1}$.
Unfortunately, all but one of the galaxies in this diagram have a low accuracy
of distance measurements.

At the bottom plot of Fig.$\;$4 the effect of galaxy outflows from the Local
Void is noticeable only as a trend. The median radial velocity follows on the
average by 10--15\% higher than the global value of $H_0$ =
73~(km$\:$s$^{-1}$)/Mpc, which is consistent with theoretical expectations. One
galaxy with a high-precision distance of 7.8$\pm$0.6 Mpc (KK~246=ESO~461-36) is
located within the void at its eastern near side and has a negative peculiar
velocity to us of about 110~km$\:$s$^{-1}$.

It must be emphasized that the Hubble diagrams in Fig.$\;$4 do not take into
account the complete observational data on the kinematics of galaxies,
surrounding the void. For this reason, we used also another approach: building a
Hubble diagram, but for velocities and distances of galaxies relative to the
center of the Local Void. At the distance of a galaxy from the observer $R_g$,
and the distance of the void's center from the observer $R_c$, the squared
distance of the galaxy from the void center is
$$R^2_{LV}=R^2_g+R^2_c-2R_gR_c\cos\Theta.$$

However, in contrast to the distance, a transition from the radial velocity of
the galaxy to its velocity relative to the void center is ambiguous as we do not
know the complete velocity vector of the galaxy, namely, its tangential
component. We viewed this issue on an example of the Virgo-centric flow
(Karachentsev et al. 2010). In the case when galaxies are involved in a weakly
perturbed Hubble flow (a model of a small void), their velocity relative to the
center of the void is expressed as

$$V_{LV}=V_g \cos{\lambda}-V_c \cos(\lambda+\Theta),$$ where $V_g$ and $V_c$
denote the radial velocities of the galaxy and the void center, and $\lambda$ is
the angle between the line of sight and the line connecting the galaxy with the
void center:

$$\tan\lambda= R_c \sin\Theta/(R_g-R_c \cos\Theta).$$

In another extreme case, when the dominant motion of galaxies is their radial
recession from the void (a model of extended void), the velocity of galaxies
relative to the void center is expressed as $$V_{LV}=(V_g -V_c\cos\Theta)
\sec\lambda$$ (see Fig.$\;$2 in Karachentsev et al. 2010). The differences between
the two assumed schemes of galaxy motions become significant at the angles
$45^{\circ}<\lambda<135^{\circ}$.

\begin{figure}[b]
\begin{center}
\includegraphics[height=0.8\textwidth,keepaspectratio,angle=270]{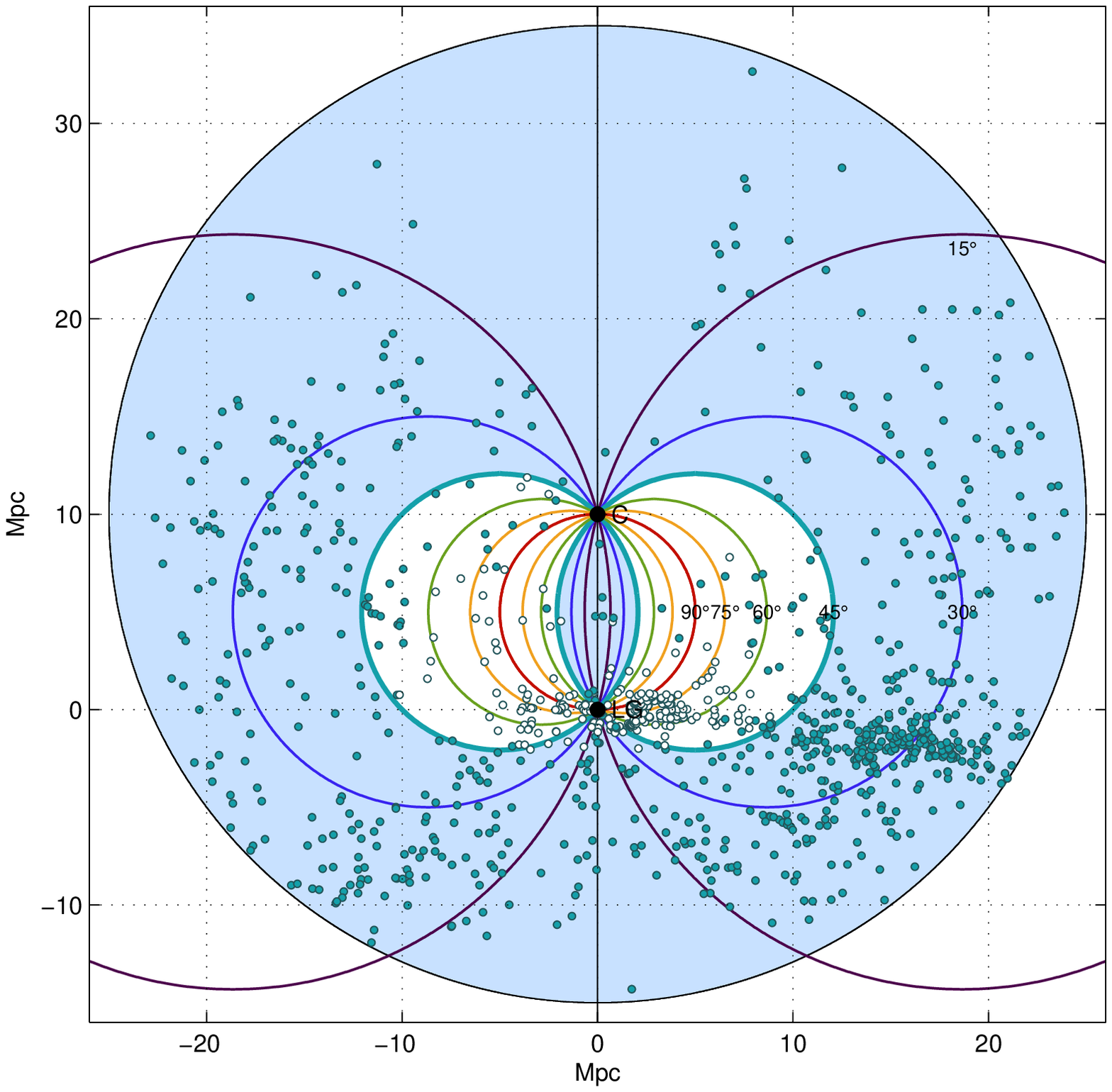}
\vbox{\vspace{0.4cm}
\parbox{0.8\textwidth}{\footnotesize{}%
Fig.$\;$5. Distribution of galaxies around the Local Void projected on the plane
passing through the Local Group (LG) and the Void center (C). The circles
correspond to the lines on which the angle $\lambda$ between the line of sight
and the direction from the galaxy towards the center of the void takes the
values: $15^{\circ}, 30^{\circ},\ldots 90^{\circ}$. The galaxies, located in the
zone of ``unfavorable ''values of $45^{\circ}<\lambda<135^{\circ}$ are shown
with empty circles.}}
\end{center}
\end{figure}

\enlargethispage{-\baselineskip}

Figure~5 illustrates the situation, showing the distribution of 1056 galaxies
with known velocities and distances within a radius of $R_{VC}=25$ Mpc around
the center of the Local Void in Cartesian coordinates. The distribution of these
galaxies is projected onto the plane passing through the Local Group (LG) and
the Local Void center (C). The selected projection plane approximately
corresponds to the Supergalactic plane \{SGZ, SGY\}, where the concentration of
galaxies in the lower right part exists owing to the Virgo cluster. The lines of
the fixed angle $\lambda= 15^{\circ}, 30^{\circ},\ldots 90^{\circ}$ are marked
in the Figure with symmetrical circles. The galaxies with angles
$45^{\circ}<\lambda<135^{\circ}$ are characterized by the greatest uncertainty
of transition from $V_g$ to $V_{LV}$ due to an unknown tangential component of
their velocity relative to the observer. These ``bad'' galaxies are shown with
open circles. In space the region $45^{\circ}<\lambda<135^{\circ}$ has the form
of a torus, the projection of which in Fig.$\;$5 looks like a white apple.

\begin{figure}[p]
\begin{center}
\includegraphics[height=0.8\textwidth,keepaspectratio,angle=270]{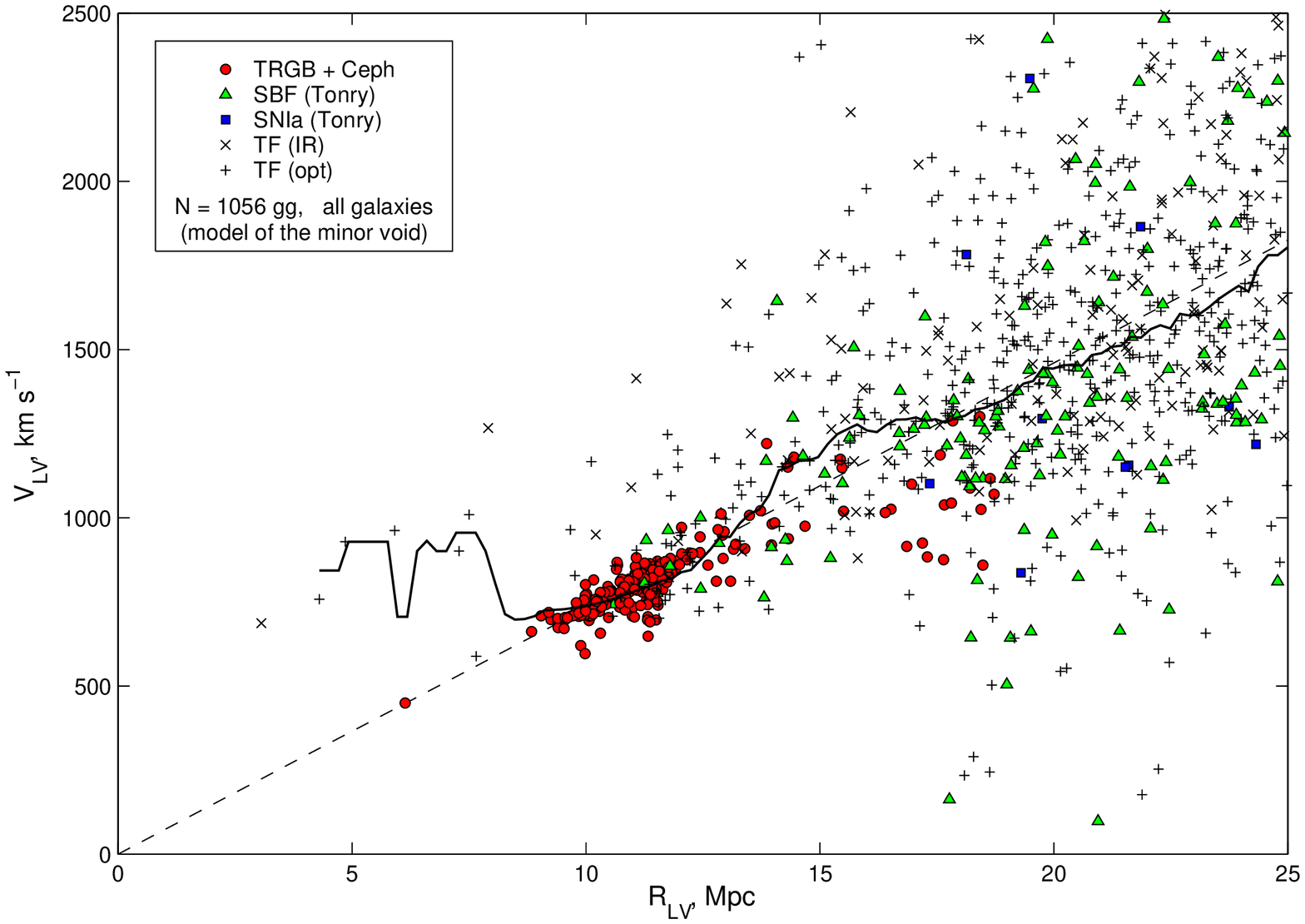}
\includegraphics[height=0.8\textwidth,keepaspectratio,angle=270]{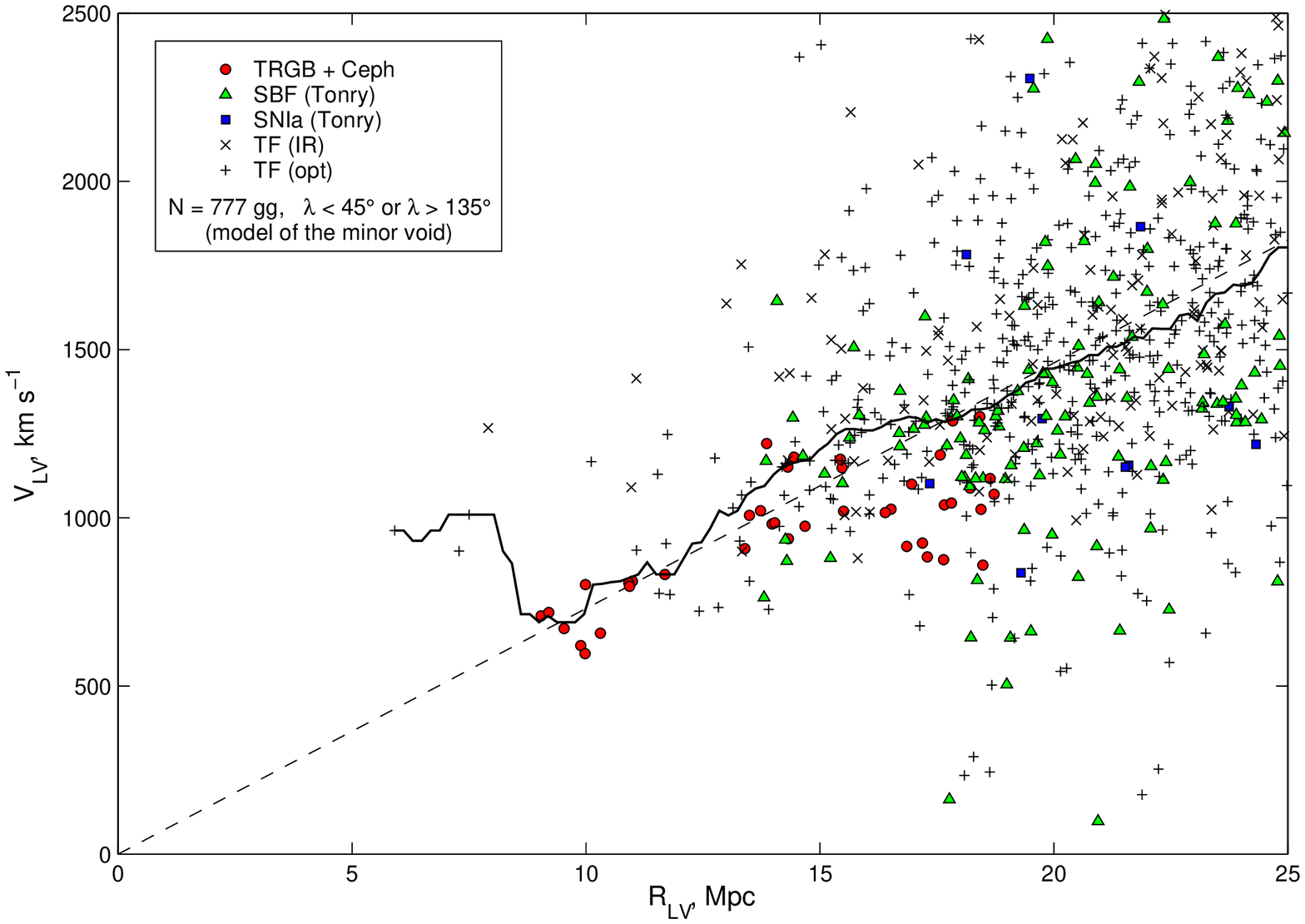}
\vbox{\vspace{0.4cm}
\parbox{0.8\textwidth}{\footnotesize{}%
Fig.$\;$6. The relationship between the velocities and distances of galaxies
relative to the center of the Local Void. The subsamples of galaxies with
distances estimated via different methods are marked with the same symbols as in
Fig.3. The top plot represents all the galaxies, the bottom plot --- the
galaxies with a ``favorable'' orientation angles $\lambda$.}}
\end{center}
\end{figure}

The distribution of galaxies by velocity and distance from the center of the
Local Void is presented in Fig.$\;$6. The upper panel represents the
distribution of all the 1056 galaxies, while the bottom plot shows the
\{$V_{LV},R_{LV}$\} distribution only for 777 galaxies with corresponding angles
of $\lambda<45^{\circ}$ or $\lambda>135^{\circ}$. The dashed line in the panels
corresponds to the unperturbed Hubble flow with the parameter
$H_0=73$~(km$\:$s$^{-1}$)/Mpc, and the broken solid line represents the sliding
median with a window of 2 Mpc. In both panels outside $R_{LV}$ =10 Mpc, the
sliding median reasonably follows the linear Hubble relation, and at $R_{LV}<10$
Mpc it reveals a velocity excess of $\sim300$~km$\:$s$^{-1}$, which clearly
indicates an accelerated expansion of the void's edges. While constructing these
diagrams the observed radial velocities of galaxies were converted into the
velocities $V_{LV}$ based on the model of a small void. The use of another
assumption (the model of extended void) yields roughly the same picture for 777
galaxies with angles $\lambda<45^{\circ}$ or $\lambda>135^{\circ}$.

\section{Galaxies within the Local Void}

\begin{figure}[h]
\begin{center}
\includegraphics[width=7.8cm,keepaspectratio]{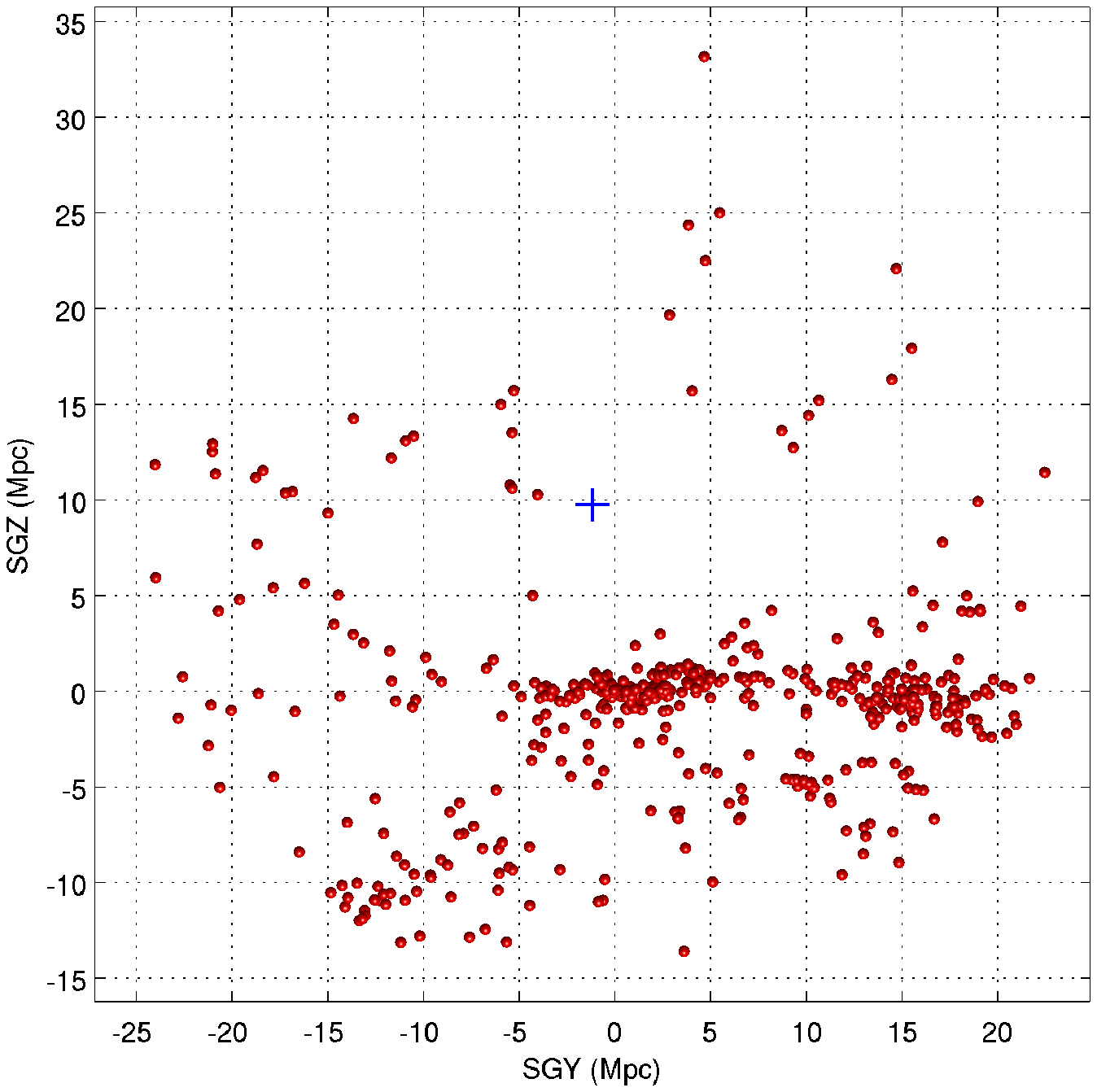}\hspace{0.2cm}%
\includegraphics[width=7.9cm,keepaspectratio]{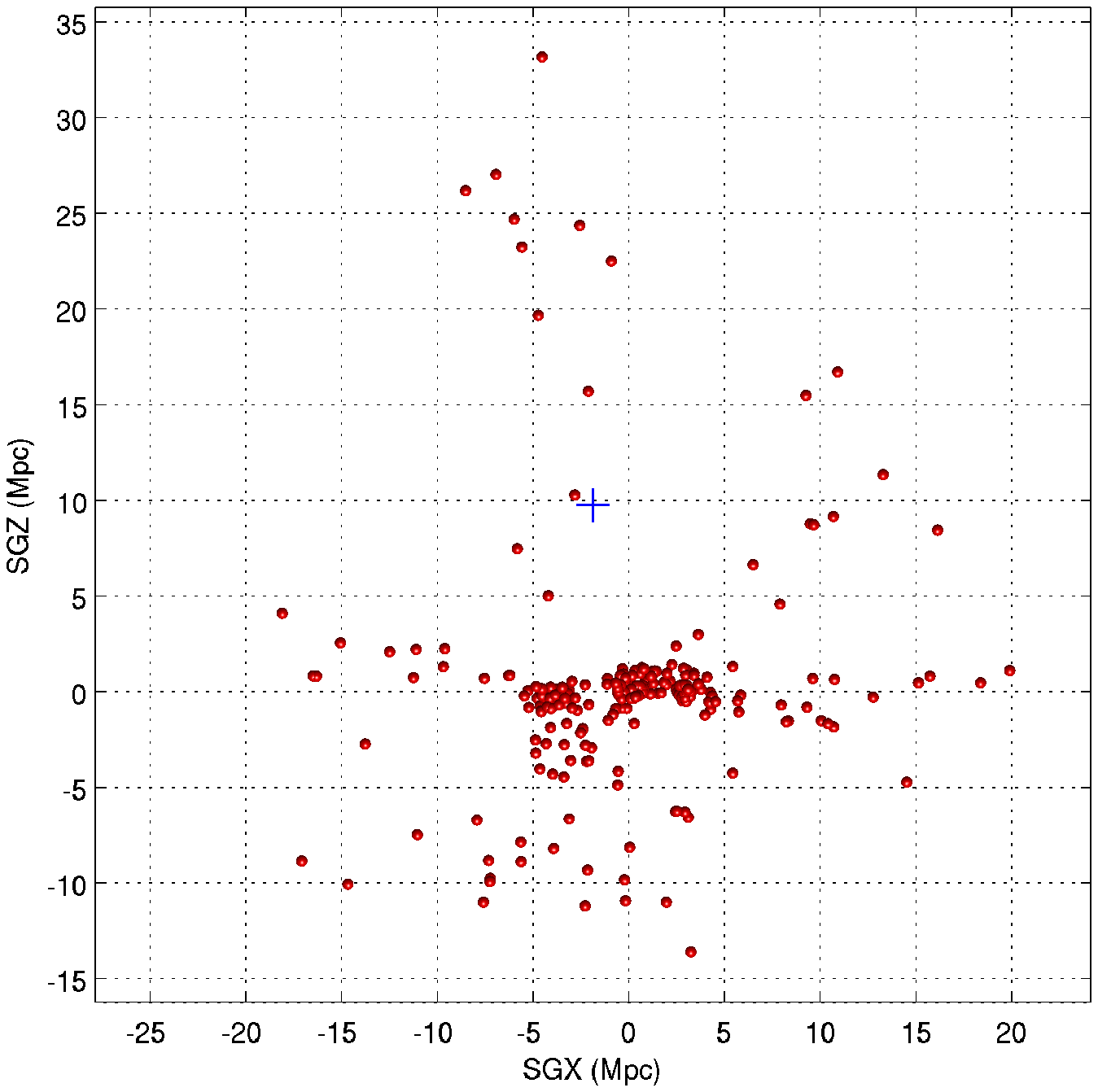}
\vbox{\vspace{0.4cm}
\parbox{0.8\textwidth}{\footnotesize{}%
Fig.$\;$7. Distribution of galaxies around the Local Void in projection on the
Supergalactic \{SGZ, SGY\} and \{SGZ, SGX\} planes. The position of the void's
center is indicated by the cross. The left and right panels demonstrate only the
galaxies within the layer $\mid$SGX$\mid <5$ Mpc and $\mid$SGY$\mid <5$ Mpc,
respectively.}}
\end{center}
\end{figure}

Current ideas on the structure of empty cosmic volumes admit the existence
inside of them of a small number of galaxies, generally of low luminosity
(Peebles 2001, Patiri et al. 2006, Hoeft \& Gottloeber 2010). An observational
verification of this hypothesis turns out to be rather difficult, and can be
successful only for the closest voids. Distribution of the considered galaxies
in the vicinity of the Local Void is presented in the plots in Fig.$\;$7 being
projected in the Supergalactic \{SGZ, SGY\} and \{SGZ, SGX\} planes. To
highlight the contours of the Void, the plots contain the galaxies within $\pm5$
Mpc layers from the respective planes. The contrast of the Local Void is notable
here, however, it is expressed most distinctly if the line of sight is turned by
the angle $\sim45^{\circ}$ relative to SGX.

With the distance estimates from the Local Void center for all the galaxies, we
determined the variation in the number density of galaxies along the radius of
the Void. We used their mean number density in the sphere of radius $R_{LV}=25$
Mpc as normalization. The resulting profile of the Local Void is presented in
Fig.$\;$8. Here the solid circles outline the density profile with the position
of the Void center at the point \{RA=19.0$^h$, Dec=$+3^{\circ}$\}, and the empty
circles correspond to the center position at \{RA=18.6$^h$, Dec=$+18^{\circ}$\}.
As we see, general shape of the $n(R_{LV})/\bar{n}$ profile changes little when
the void center is shifted. Inside the void, the mean number density of galaxies
amounts to about 1/5 of the average density. Directly at the edge of the void at
$R_{LV}\simeq(10-12)$ Mpc there is a peak density, which is obviously formed by
the filaments imbordering the void. It should be noted that the nearby half of
the Local Void is practically free from galaxies down to the absolute magnitude
$M_{B}\simeq-10^m$, whereas the far half of the void reveals galaxies, the
distances to which are measured with a $\sim20$\% error, i.e. $\sim(3-4)$ Mpc.
Therefore, the reality of galaxies being present within the void needs further
confirmations with more accurate measurements of their distances.

\begin{figure}[h]
\begin{center}
\includegraphics[height=0.8\textwidth,keepaspectratio,angle=270]{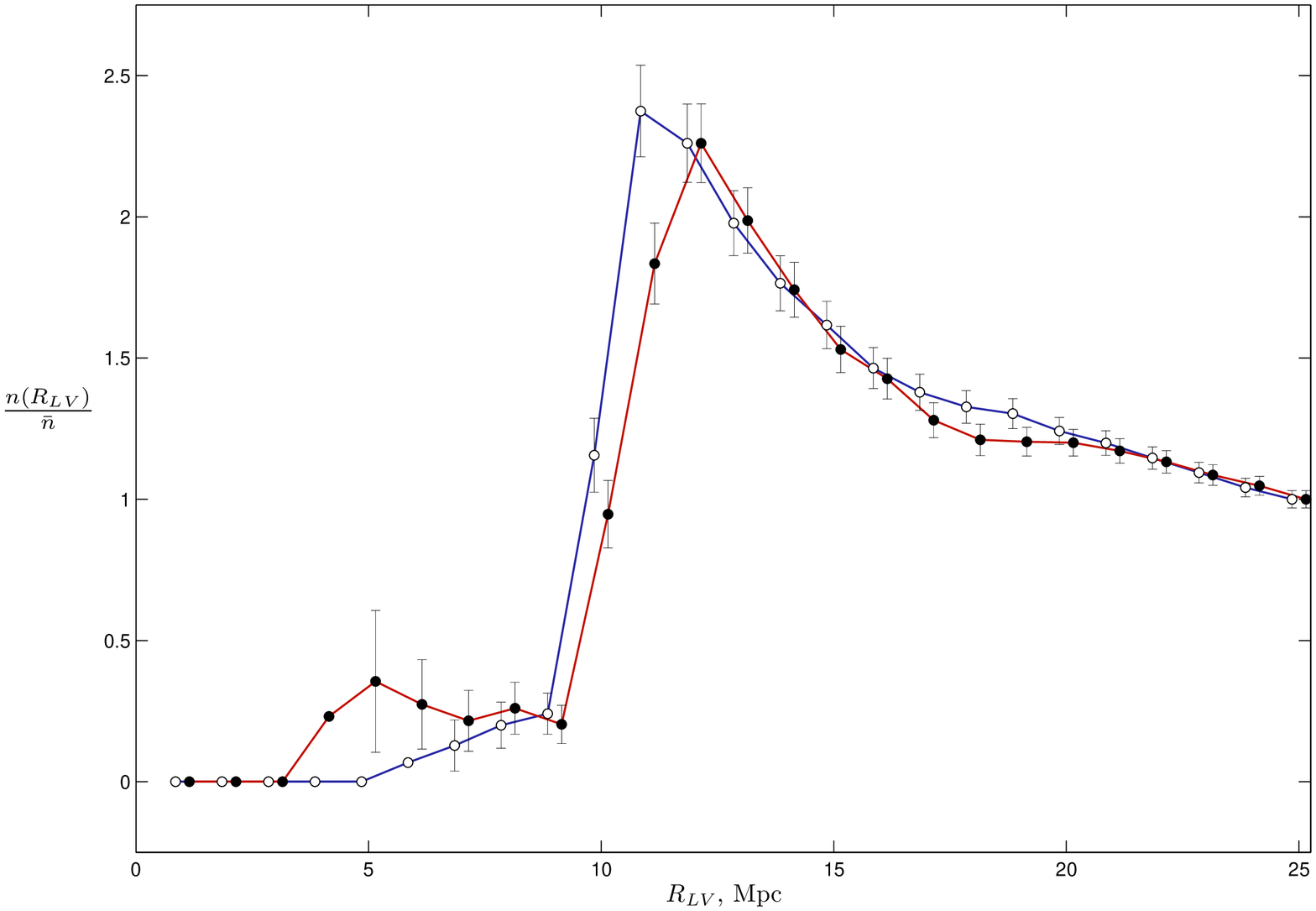}
\vbox{\vspace{0.4cm}
\parbox{0.8\textwidth}{\footnotesize{}%
Fig.$\;$8. Density number profile of galaxies along the radius of the Local Void
at two positions of its center.}}
\end{center}
\end{figure}

Some of the data on 16 galaxies within the Local Void are presented in Table 2.
In addition to the distances and velocities relative to the Local Group, the
columns contain: the distances from the void center $R_{LV}$ (in Mpc),
velocities relative to the center of the void (in km$\:$s$^{-1}$) when
calculated according to the model of a small and extended void, as well as the
morphological types and absolute magnitudes of the galaxies. The galaxies are
ranked by the distance value $R_{LV}$. The column ``$\lambda$'' lists the values
of the angle, characterizing the galaxy position relative to the line of sight
and the direction towards the center of the void.

\begin{table}
\caption{\footnotesize{}Galaxies nearest to the Local Void center
(RA=19.0$^h$, Dec=+3$^{\circ}$)}
\begin{center}
{\footnotesize{}%
\begin{tabular}{lcrrlrrrr} \hline
 Galaxy     &  RA(2000.0)Dec  & $R_{LG}\pm\sigma$&$V_{LG}$&$R_{LV}$ & $V_{LV}$& T &$M_B$ &$\lambda$\\
\hline
2MFGC15085  & 194311.7-065621 & 11.4$\pm$2.3 &1654 & 3.06 &  687/ 1683 &  8 & -17.2&  56  \\
UGCA417     & 200921.2-061710 & 12.1$\pm$2.4 &1595 & 4.30 &  758/ 1455 &  8 & -15.5&  51  \\
NGC6835     & 195433.1-123409 & 12.7$\pm$2.5 &1761 & 4.86 &  928/ 1567 &  6 & -18.1&  47  \\
NGC6821     & 194424.1-064960 & 15.0$\pm$3.0 &1683 & 5.91 &  962/ 1084 &  8 & -17.1&  26  \\
KK246       & 200357.4-314054 &  7.8$\pm$0.6 & 458 & 6.14 &  449/11257 & 10 & -13.7&  91  \\
CGCG371-004 & 194326.0-011032 & 16.8$\pm$3.4 &1610 & 7.29 &  901/  931 &  5 & -18.7&  16  \\
CGMW 3-4603 & 194238.0-073639 & 16.7$\pm$3.3 &1711 & 7.50 & 1009/ 1072 &  8 & -17.0&  20  \\
MCG-01-41-6 & 160936.8-043713 & 10.7$\pm$2.1 & 847 & 7.65 &  588/  706 &  9 & -15.6&  64  \\
2MFGC14044  & 174804.5+144429 & 16.3$\pm$3.3 &1969 & 7.91 & 1267/ 1450 &  8 & -17.7&  27  \\
N6789       & 191641.8+635822 &  3.6$\pm$0.3 & 144 & 8.84 &  662/ 1491 &  9 & -14.3&  98  \\
SagDIG      & 192959.0-174041 &  1.0$\pm$0.1 &  23 & 9.04 &  708/  718 & 10 & -11.5& 156  \\
DDO210      & 204651.8-125053 &  0.9$\pm$0.1 &  13 & 9.21 &  718/  739 & 10 & -11.1& 146  \\
KKR25       & 161347.6+542216 &  1.9$\pm$0.2 &  68 & 9.26 &  697/  891 & -1 & - 9.9& 108  \\
IC 4662     & 174706.3-643825 &  2.4$\pm$0.2&  145&  9.40&   691/  957 &  9 & -15.6&  97  \\
IC 5152     & 220241.9-511743 &  2.0$\pm$0.2 &  75 & 9.40 &  700/ 1012 &  9 & -15.7& 102  \\
N6503       & 174927.6+700841 &  5.3$\pm$0.4 & 301 & 9.41 &  673/  177 &  6 & -18.1&  81  \\
\hline
\end{tabular}}
\end{center}
\end{table}

It follows from the data above that the sparse population of the Local Void
mostly contains the galaxies of late types and low luminosity. The median
absolute magnitude of galaxies in Table 2 is $-15.7^m$, and the median
morphological type corresponds to $T=8$ or Sdm. The only early-type galaxy in
the Table, KKR~25, belongs to the rare type of isolated dwarf spheroidal
galaxies. Due to their very low luminosity ($M_B=-9.9^m$), such objects are
difficult to detect, but can be quite common in the space between groups and
clusters (Karachentsev et al. 2009).

\section{Concluding remarks}

We did not find any detailed data in the literature on the structure and
kinematics of a typical cosmic void. Clearly, the numerical simulations of the
process of void's expansion under the $\Lambda$CDM paradigm allow one to draw in
Figures 4 and 6 the regression lines, which give a quantitative description of
the expansion pattern. According to Tully et al. (2008), the expansion of
absolutely empty spherically symmetric void at $\Omega_{\lambda}=0.76$ is
characterized by the local Hubble constant $H_v\simeq
H_0+16$~(km$\:$s$^{-1}$)/Mpc. Then, at the diameter of the Local Void we adopted
as 20 Mpc, the amplitude of outflow velocity differential amounts to
$\sim320$~km$\:$s$^{-1}$. This value is quite consistent with the available
observational data. Of course, the kinematics of a real nonspherical void may
differ significantly from this simplest model.

It should be noted that we have not yet exhausted the observational capabilities
to clarify the expansion pattern of the Local Void. In its near side there are
some galaxies, the distances to which can be measured using the TRGB method with
an accuracy better than 5\%. At the far edge of the void, there remain a lot of
spiral galaxies that are seen nearly edge-on, and for which so far there are no
accurate measurements of the HI line profiles, and no reliable photometry. With
the view of the current GBT monitoring such kind galaxies (Courtois et al.
2009), and the photometric PanStarr survey, new and more reliable observational
data on the kinematics of the Local Void may appear in the near future.

\bigskip

{\bf Acknowledgements}

\noindent{}Authors thank Brent Tully for useful discussions and valuable comments.
\vspace{-0.3cm}
\begin{tabbing}
We made use of \=the\hspace{0.2cm}\=HyperLEDA (http://leda.univ-lyon1.fr/)\\
\>and \>NED (http://nedwww.ipac.caltech.edu/) databases.\\
\end{tabbing}

\vspace{-1.3cm}
\begin{tabbing}
This work was supported by the grants: \=RFBR 09--02--90414-Ukr-f-a and\\
\>RFBR 10--02--92650. \\
\end{tabbing}

\section*{References}
\pagebreak[3]\hangindent=1cm\noindent1. {\itshape{}Ceccarelli L., Padilla N.\,D., Valotto C., Lambas D.\,G., }2008, arXiv:0805.0797\par
\pagebreak[3]\hangindent=1cm\noindent2. {\itshape{}Courtois H., Tully B., Fisher R. et al. }2009, AJ, {\bfseries{}138}, 1938\par
\pagebreak[3]\hangindent=1cm\noindent3. {\itshape{}de Lapparent V., Geller M.\,J., Huchra J.\,P., }1986, ApJ, {\bfseries{}302}, L1\par
\pagebreak[3]\hangindent=1cm\noindent4. {\itshape{}Faber S.\,M., Burstein D., }1988, in ``Large-Scale Motions in the Universe'', A Vatican study Week, p.115\par
\pagebreak[3]\hangindent=1cm\noindent5. {\itshape{}Giovanelli R., et al., }2005, AJ, {\bfseries{}130}, 2598\par
\pagebreak[3]\hangindent=1cm\noindent6. {\itshape{}Gregory S.\,A., Thompson L.\,A., }1978, ApJ, {\bfseries{}222}, 784\par
\pagebreak[3]\hangindent=1cm\noindent7. {\itshape{}Hoeft M., Gottloeber S., }2010, arXiv:1001.4721\par
\pagebreak[3]\hangindent=1cm\noindent8. {\itshape{}Joeveer M., Einasto J., Tago E., }1978, MNRAS, {\bfseries{}185}, 357\par
\pagebreak[3]\hangindent=1cm\noindent9. {\itshape{}Karachentsev I.\,D., Nasonova (Kashibadze) O.\,G., }2010, MNRAS, {\bfseries{}405}, 1075\par
\pagebreak[3]\hangindent=1cm\noindent10. {\itshape{}Karachentsev I.\,D., Kashibadze O.\,G., Makarov D.\,I., Tully R.\,B., }2009, MNRAS, {\bfseries{}393}, 1265\par
\pagebreak[3]\hangindent=1cm\noindent11. {\itshape{}Karachentsev I.\,D., Dolphin A.\,E., Tully R.\,B., et al., }2006, AJ, {\bfseries{}131}, 1361\par
\pagebreak[3]\hangindent=1cm\noindent12. {\itshape{}Karachentsev I.\,D., Karachentseva V.\,E., Huchtmeier W.\,K., Makarov D.\,I., }2004, AJ, {\bfseries{}127}, 2031 (= CNG)\par
\pagebreak[3]\hangindent=1cm\noindent13. {\itshape{}Karachentseva V.\,E., Karachentsev I.\,D., Richter G.\,M., }1998, A\&AS, {\bfseries{}134}, 1\par
\pagebreak[3]\hangindent=1cm\noindent14. {\itshape{}Kashibadze O.\,G., }2008, Astrofizika, {\bfseries{}51}, 409\par
\pagebreak[3]\hangindent=1cm\noindent15. {\itshape{}Kraan-Korteweg R.\,C., Shafi N., Koribalski B.\,S., et al., }2008, in ``Galaxies in the Local Volume'', ASSP, p.13\par
\pagebreak[3]\hangindent=1cm\noindent16. {\itshape{}Lee M.\,G., Freedman W.\,L. Madore B.\,F., }1993, ApJ, {\bfseries{}417}, 553\par
\pagebreak[3]\hangindent=1cm\noindent17. {\itshape{}Mitronova S.\,N., Karachentsev I.\,D., Karachentseva V.\,E., Jarrett T.\,H., Kudrya Yu.\,N., }2004, BSAO, {\bfseries{}57}, 5 (2MFGC)\par
\pagebreak[3]\hangindent=1cm\noindent18. {\itshape{}Nakanishi K., Takata T., Yamada T., et al., }1997, ApJS, {\bfseries{}112}, 245\par
\pagebreak[3]\hangindent=1cm\noindent19. {\itshape{}Nasonova O.\,G., de Freitas Pacheco J.\,A., Karachentsev I.\,D., }A\&A, submitted\par
\pagebreak[3]\hangindent=1cm\noindent20. {\itshape{}Patiri S.\,G., Betancort-Rijo J., Prada F., et al., }2006, MNRAS, {\bfseries{}372}, 1710\par
\pagebreak[3]\hangindent=1cm\noindent21. {\itshape{}Peebles P.\,J.\,E., }2001, ApJ, {\bfseries{}557}, 495\par
\pagebreak[3]\hangindent=1cm\noindent22. {\itshape{}Roman A.\,T., Nakanishi K., Tomita A., Saito M., }1996, PASJ, {\bfseries{}48}, 679\par
\pagebreak[3]\hangindent=1cm\noindent23. {\itshape{}Saintonge A., Giovanelli R., Haynes M.P., et al., }2008, AJ, {\bfseries{}135}, 588\par
\pagebreak[3]\hangindent=1cm\noindent24. {\itshape{}Schaap W., }2007, PhD Thesis, Groningen Univ.\par
\pagebreak[3]\hangindent=1cm\noindent25. {\itshape{}Schlegel D.\,J., Finkbeiner D.\,P. \& Davis M., }1998, ApJ, {\bfseries{}500}, 525\par
\pagebreak[3]\hangindent=1cm\noindent26. {\itshape{}Springob C.\,M., Haynes M.\,P., Giovanelli R., Kent B.\,R., }2005, ApJS, {\bfseries{}160}, 149\par
\pagebreak[3]\hangindent=1cm\noindent27. {\itshape{}Tikhonov A.\,V., Karachentsev I.\,D., }2006, ApJ, {\bfseries{}653}, 969\par
\pagebreak[3]\hangindent=1cm\noindent28. {\itshape{}Tonry J.\,L., Schmidt B.\,P., Barris B., et al., }2003, ApJ, {\bfseries{}594}, 1\par
\pagebreak[3]\hangindent=1cm\noindent29. {\itshape{}Tonry J.\,L., Dressler A., Blakeslee J.\,P., et al., }2001, ApJ, {\bfseries{}546}, 681\par
\pagebreak[3]\hangindent=1cm\noindent30. {\itshape{}Tully R.\,B., Rizzi L., Shaya E.\,J., Courtois H.\,M., Makarov D.\,I., Jacobs B.\,A., }2009, AJ, {\bfseries{}138}, 323\par
\pagebreak[3]\hangindent=1cm\noindent31. {\itshape{}Tully R.\,B., Shaya E.\,J., Karachentsev I.\,D., et al., }2008, ApJ, {\bfseries{}676}, 184\par
\pagebreak[3]\hangindent=1cm\noindent32. {\itshape{}Tully R.\,B., Rizzi L., Dolphin A.\,E., et al., }2006, AJ, {\bfseries{}132}, 729\par
\pagebreak[3]\hangindent=1cm\noindent33. {\itshape{}Tully R.\,B., Fisher J.\,R., }1987, Nearby Galaxies Atlas, Cambridge Univ. Press, Cambridge\par
\pagebreak[3]\hangindent=1cm\noindent34. {\itshape{}van de Weygaert R., van Kampen E., }1993, MNRAS, {\bfseries{}263}, 481\par
\pagebreak[3]\hangindent=1cm\noindent35. {\itshape{}Zwaan M.A., Staveley-Smith L., Koribalski B.S., et al., }2003, AJ, {\bfseries{}125}, 2842\par

\end{document}